\documentstyle[preprint,tighten,aps,floats,psfig,epsfig]{revtex}

\begin{document}
\draft
\title{
Crossover from random-exchange to random-field critical behavior
in Ising models.
}
\author{Pasquale Calabrese,$^1$ Andrea Pelissetto,$^2$ 
Ettore Vicari$^3$ }
\address{$^1$ Scuola Normale Superiore and  INFN, Piazza dei Cavalieri 7,
 I-56126 Pisa, Italy.}
\address{$^2$ Dipartimento di Fisica dell'Universit\`a di Roma I
and INFN, I-00185 Roma, Italy.}
\address{$^3$
Dipartimento di Fisica dell'Universit\`a 
and INFN, 
Via Buonarroti 2, I-56127 Pisa, Italy.
{\bf e-mail: \rm 
{\tt calabres@df.unipi.it},
{\tt Andrea.Pelissetto@roma1.infn.it},
{\tt vicari@df.unipi.it}.
}}

\date{\today}

\maketitle

\begin{abstract}
We compute the crossover exponent $\phi$ describing the
crossover from the random-exchange to the random-field
critical behavior in Ising systems.
For this purpose, we consider the field-theoretical
approach based on the replica method,
and perform a six-loop calculation
in the framework of a fixed-dimension expansion.
The crossover from random-exchange to random-field
critical behavior has been
observed in dilute anisotropic
antiferromagnets, such as Fe$_x$Zn$_{1-x}$F$_2$
and Mn$_x$Zn$_{1-x}$F$_2$,
when applying an external magnetic field.
Our result $\phi=1.42(2)$ for the crossover exponent 
is in good agreement with the available experimental estimates.
\end{abstract}

\pacs{PACS Numbers: 75.10.Nr, 75.10.Hk, 64.60.Ak}


The most studied experimental realizations of random-field
Ising systems  are dilute anisotropic antiferromagnets
in a uniform magnetic field applied along the 
spin ordering axis \cite{FA-79,WMD-82,Cardy-84}.
A simple lattice model is provided by the Hamiltonian
\begin{equation}
{\cal H} = J \sum_{<ij>}  \rho_i \,\rho_j \; s_i s_j
- H \sum_i \rho_i s_i,
\label{Hamiltonian}
\end{equation}
where $J>0$, the first sum extends over all nearest-neighbor sites,
$s_i=\pm 1$ are the spin variables, and
$\rho_i$ are uncorrelated quenched random variables, which are equal to one 
with probability $p$ (the spin concentration) and zero with probability $1-p$
(the impurity concentration). 
In the absence of an external field, i.e. $H=0$, 
the critical behavior of dilute Ising systems 
(above the percolation point of the spins) belongs to the
universality class of the random-exchange Ising model (REIM),
which differs from the standard Ising one.
See, e.g., Refs. \cite{PV-r,FHY-01} for recent reviews. 
The applied uniform field $H$ gives rise to 
a different critical behavior corresponding to the universality class of
the random-field Ising  model (RFIM).
See e.g. Refs.~\cite{Belanger-98,Nattermann-98,Belanger-00}
for recent reviews on RFIM.

For $t\equiv(T - T_N)/T_N\to 0$, $T_N$ being the N\'eel temperature in
zero field, and $H\to 0$ the  singular part of the free energy can be 
written as \cite{FA-79}
\begin{equation}
{\cal F} = |u_t|^{2-\alpha} f\left( H^2 |u_t|^{-\phi} \right),
\label{screl}
\end{equation}
where $u_t\approx t + a_1 H^2 + a_2 t^2 $ is the scaling field associated with
the temperature, $\alpha$ is the specific-heat exponent of the REIM,
$f(x)$ is a scaling function,
and $\phi$ is the corresponding crossover exponent.
As a consequence of the crossover scaling (\ref{screl}), 
the critical temperature in the presence of the external 
field $H$ is given by
\begin{equation}
T_c(H) \approx T_N + c H^{2/\phi} - a H^2,
\label{Tch}
\end{equation}
for sufficiently small $H$.
Experimental measurements on dilute antiferromagnets yielded  
rather precise estimates:
$\phi=1.42(3)$ obtained for Fe$_x$Zn$_{1-x}$F$_2$ \cite{FKJ-91},
$\phi=1.43(3)$ for Mn$_x$Zn$_{1-x}$F$_2$ \cite{RKJ-88},
and 
$\phi=1.41(5)$ for Fe$_x$Mg$_{1-x}$Cl$_2$ \cite{LK-84}.

On the theoretical side, a computation of $\phi$ was
presented in Ref.~\cite{Aharony-86} using
the field-theoretical $\epsilon$ expansion.
We recall that in the REIM  the $\epsilon$ expansion 
is actually an expansion in powers of $\sqrt{\epsilon}$.
The exponent $\phi$ was computed to $O(\epsilon)$ \cite{Aharony-86},
obtaining
\begin{equation}
\phi/\gamma=1 + 0.16823 \epsilon^{1/2} 
- 0.22666 \epsilon + O(\epsilon^{3/2}),
\end{equation}
where $\gamma$ is the REIM susceptibility exponent. 
Apart from showing that $\phi\neq \gamma$, 
where $\gamma\approx 1.34$ 
(see, e.g., Refs.~\cite{CMPV-03,BFMMPR-98,PV-00}),
this series does not yield a reliable estimate of $\phi$
for three-dimensional systems.
This is generically true for the $\sqrt{\epsilon}$ expansion
of any critical exponent of the REIM \cite{PV-r,FHY-01}.

In this paper we determine the crossover exponent $\phi$
using an alternative field-theoretical method 
based on a fixed-dimension expansion
in powers of appropriate zero-momentum quartic couplings,
which we compute to six loops.
As we shall see, the analysis of the series provides
a quite precise estimate of $\phi$,
\begin{equation}
\phi=1.42(2),
\label{final}
\end{equation} 
in good agreement with experiments.
This result was already anticipated in Ref.~\cite{th2002}.

The field-theoretical approach is based on 
an effective Landau-Ginzburg-Wilson $\varphi^4$ Hamiltonian
that can be obtained by using the replica method \cite{replica}, i.e.
\begin{equation}
{\cal H}_{\varphi^4} =  \int d^3 x \Bigl\{ {1\over 2} \sum_{i=1}^{N}
      \left[ (\partial_\mu \varphi_i)^2 +  r \varphi_i^2 \right]  
+{1\over 4!} \sum_{i,j=1}^N \left( u_0 + v_0 \delta_{ij} \right)
\varphi^2_i \varphi^2_j  \Bigr\},
\label{Hphi4rim}
\end{equation}
where $\varphi_i$ is an $N$-component field.
The critical behavior of the REIM is expected to be described
by the stable fixed point of the 
Hamiltonian ${\cal H}_{\varphi^4}$ in the limit $N\rightarrow 0$ 
for $u_0<0$ \cite{CPPV-03}.
The most precise field-theoretical results for the critical exponents
have been obtained by analyzing the fixed-dimension 
expansion in powers of the
zero-momentum quartic couplings $u,v$ related to $u_0,v_0$
(i.e. $u = Z_u u_0/m$ and $v=Z_v v_0/m$ where 
$Z_{u,v}=1 + O(u,v)$), which have been computed to six 
loops \cite{PV-00,CPV-00}.
We refer the reader to Ref.~\cite{PV-00} for notations
and definitions.

In the presence of a spatially uncorrelated
random field $h(x)$ with zero average and variance $h^2$,
one may still apply the replica method.
Averaging the replicated random field term 
\begin{equation}
\int d^3 x \,h(x) \sum_i \varphi_i(x)
\end{equation}
over the Gaussian distribution of the field,
one obtains a new term proportional to  
\begin{equation}
h^2 \int d^3 x\,\sum_{i,j} \varphi_i(x)\varphi_j(x),
\end{equation}
that must be added to the
Hamiltonian ${\cal H}_{\varphi^4}$ \cite{Aharony-86}.
This term causes the crossover from the REIM to the RFIM. Thus,
the REIM-to-RFIM crossover exponent is determined by the
$N\rightarrow 0$ limit of the 
renormalization-group (RG) dimension $y_T$ of the quadratic 
operator~\cite{footnoteU}
\begin{equation}
T_{ij} = \varphi_i \varphi_j, \qquad i \neq j.
\label{Tdef}
\end{equation}
In order to evaluate $y_T$ and the corresponding
crossover exponent $\phi=y_T\nu$, we define a related
RG function $Z_T$ from the one-particle irreducible
two-point function $\Gamma_T^{(2)}$ with an insertion of the operator
$T_{ij}$, i.e.
\begin{equation}
\Gamma_T^{(2)}(0)_{ij,kl} = Z_T^{-1} \; A_{ijkl},
\end{equation}
where
$A_{ijkl} = \delta_{ik}\delta_{jl} + \delta_{il}\delta_{jk}$
so that $Z_T(0)=1$.
Then, we compute  the RG function 
\begin{equation}
\eta_T(u,v) = \left. {\partial \ln Z_T \over \partial \ln m}\right|_{u_0,v_0}
= \beta_u {\partial \ln Z_T \over \partial u} +
\beta_v {\partial \ln Z_T \over \partial v} ,
\end{equation}
where $\beta_u$ and $\beta_v$ are the $\beta$-functions.
The exponent $\eta_T$ is given by the value of $\eta_T(u,v)$ 
at $u=u^*$, $v=v^*$, where ($u^*$,$v^*$) is the REIM stable fixed point.
Finally, the REIM-to-RFIM crossover exponent is obtained by using the
RG scaling relation
\begin{equation}
\phi = \left( 2 + \eta_T - \eta\right)\nu = \gamma + \eta_T \nu,
\label{phirg}
\end{equation}
where $\gamma$, $\nu$, and $\eta$ are  REIM critical exponents.

We computed the function $\Gamma^{(T,2)}(0)$ to six loops. 
The calculation is rather cumbersome, since it requires
the evaluation of 563 Feynman  diagrams.
We handled it with a symbolic manipulation program, which  generates the diagrams 
and computes the symmetry and group factors of each of them.
We used the numerical results compiled in Ref.~\cite{NMB-77}
for the integrals associated with each diagram.
The resulting six-loop series of $\eta_T(u,v)$ is 
\begin{eqnarray}
&&
\eta_T(\bar{u},\bar{v}) = 
-\case{1}{4} \bar{u} + \case{1}{16} \bar{u}^2 + \case{1}{18} \bar{u}\bar{v}
- 0.0357673 \,\bar{u}^3 - 0.0483240 \, \bar{u}^2 \,\bar{v} - 0.00421548 \,\bar{u}\,\bar{v}^2 
\label{seriesetat}
\\&&
+ 0.0343748 \,\bar{u}^4 + 0.0762616 \,\bar{u}^3 \,\bar{v} + 0.0416943 \,\bar{u}^2  \,\bar{v}^2
+ 0.00896116 \, \bar{u}\,\bar{v}^3 
\nonumber \\&&
- 0.0408958 \,\bar{u}^5 - 0.121377 \,\bar{u}^4  \,\bar{v} - 0.104778 \,\bar{u}^3  \,\bar{v}^2
- 0.0373712 \,\bar{u}^2  \,\bar{v}^3 - 0.00527015 \,\bar{u} \,\bar{v}^4
\nonumber \\&&
+ 0.0597048 \,\bar{u}^6 + 0.227662 \, \bar{u}^5 \,\bar{v} +  0.287108 \,\bar{u}^4\,\bar{v}^2
+ 0.17231 \,\bar{u}^3 \,\bar{v}^3 
\nonumber\\&&
+ 0.0552124 \,\bar{u}^2 \,\bar{v}^4 + 
0.00759654 \,\bar{u}\,\bar{v}^5 + ...,
\nonumber
\end{eqnarray}
where $\bar{u}$ and $\bar{v}$ are the rescaled couplings
$\bar{u} = u/(6 \pi)$ and $\bar{v} = 3 v/(16 \pi)$,
and the dots indicate higher-order terms.
In our analysis we also considered the series
$\phi(\bar{u},\bar{v})$ corresponding to $\phi$,
which can be obtained by using Eq.~(\ref{phirg}) and 
the series of the RG functions $\gamma(\bar{u},\bar{v})$ and
$\nu(\bar{u},\bar{v})$ corresponding to the critical 
exponents $\gamma$ and $\nu$, cf. Ref.~\cite{PV-00}.

In order to obtain an estimate of $\eta_T$ and $\phi$,
the corresponding six-loop series must be
resummed and then evaluated at the fixed-point values $u^*$, $v^*$.
Although the perturbative expansion is not Borel summable 
\cite{nonBorel},
various resummation schemes have been proposed and employed,
obtaining rather reliable results for the critical exponents,
see, e.g., Refs.~\cite{PV-r,FHY-01}. 
We analyzed $\eta_T(\bar{u},\bar{v})$
and $\phi(\bar{u},\bar{v})$ by using
the resummation methods outlined in Ref.~\cite{PV-00} and
the estimates $u^*=-18.6(3)$ and $v^*=43.3(2)$ obtained
by Monte Carlo simulations in Ref.~\cite{CMPV-03}, which
turn out to be more precise than those obtained from the
zeroes of the $\beta$-functions computed to six loops.

We obtained $\eta_T=0.095(30)$ from the analysis 
of $\eta_T(\bar{u},\bar{v})$, $\phi=1.43(3)$ 
from $\phi(\bar{u},\bar{v})$,
and $\phi=1.42(2)$ from $1/\phi(\bar{u},\bar{v})$.
The errors are related to the spread
of the results obtained by using different resummation methods 
and different resummation parameters within each method,
see Ref.~\cite{PV-00} for details.
Using the available estimates of the critical exponents 
(e.g., $\gamma=1.342(6)$, $\nu=0.683(3)$ and $\eta=0.035(2)$ from Monte Carlo 
simulations \cite{CMPV-03,BFMMPR-98}, 
$\gamma=1.330(17)$, $\nu=0.678(10)$ and $\eta=0.030(3)$ 
from the analysis of the six-loop
field-theoretical expansions \cite{PV-00}), the above-reported result for 
$\eta_T$ and the RG scaling relation (\ref{phirg}) give $\phi=1.41(2)$.
From these results we arrive at the
final estimate reported in Eq.~(\ref{final}),
which is in good  agreement
with the available experimental results obtained 
for various uniaxial anisotropic antiferromagnets.



\begin{references}
\bibitem{footnoteU}
In the cubic-symmetric theory (\ref{Hphi4rim})
the RG dimension of the quadratic operators
$U_i = \varphi_i^2 - {1\over N} \,\sum_a \varphi_a^2$
is different from the one of 
$T_{ij}=\varphi_i \varphi_j$ with $i\neq j$,
unlike O($N$)-symmetric theories where they 
belong to the same spin-2 irreducible representation.
In the limit $N\rightarrow 0$
the RG dimensions of $U_i$ and
the invariant operator ${1\over N} \,\sum_a \varphi_a^2$
turn out to be equal. This can be inferred from their
perturbative expansions.
\end{references}
\end{document}